\documentstyle[preprint,eqsecnum,aps,psfig]{revtex}

\begin{document}
\draft
\preprint{HEP/123-qed}
\title{Relativistic photoionization cross sections for C II}
\author{Sultana N. Nahar}
\address{ Department of Astronomy, The Ohio State University, Columbus, Ohio
43210\\ }
\date{\today}
\maketitle
\begin{abstract}
 High resolution measurements of photoionization cross sections for atomic
ions are now being made on synchrotron radiation sources.
The recent measurements by Kjeldsen etal. (1999) showed good agreement
between the observed resonance features and the
the theoretical calculations in the close coupling approximation
(Nahar 1995). However, there were several observed resonances that 
were missing in the theoretical predictions. The earlier theoretical 
calculation was carried out in $LS$ coupling where the
relativistic effects were not included. Present work reports
photoionization cross sections including the relativistic effects
in Breit-Pauli R-matrix (BPRM) approximation. The configuration 
interaction eigenfunction expansion for the core ion C III consists of
20 fine structure levels dominated by the configurations from 
$1s^22s^2$ to $1s^22s3d$. Detailed features in the calculated cross 
sections exhibit the missing resonances due to fine structure. The
present results benchmark the accuracy of BPRM photoionization cross sections 
as needed for recent and ongoing experiments.

\end{abstract}
\pacs{PACS number(s): 32.80.Fb}

\narrowtext

Although a large number of theoretical calculations for photoionization
cross sections have been carried out, the theoretical data have not yet
been benchmarked against the new generation of high-resolution 
experiments. We refer in particular the calculations using the close
coupling R-matrix method for extensively utilized in the Opacity 
Project \cite{op} and similar works that accurately considers the 
numerous autoionizing resonances in the cross sections along overlapping
Rydberg series. Most of the vast amount of photoionization data 
computed under the Opacity Project (OP), Iron Project (IP) \cite{ip}
and other works, as estimated to be accurate to 10-20\%. However,
no detailed comparisons with experiments have been possible since 
photoionization cross sections were not measured with equally detailed
features. Resonances can now be finely resolved in recent experiments
being carried out with synchrotron radiation sources in Aarhus 
\cite{kjl}, Berkely (e.g. \cite{als}), and Paris (e.g. \cite{fw}). The 
test of benchmarking the theoretical calculations with experiments is 
therefore of considerable current importance. Furthermore, the 
quality of experimental work is such as to clearly delineate the fine
structure, necessitating the inclusion of relativistic effects in an
{\it ab initio} manner. In this {\it Communication} we present and
benchmark the first theoretical results with recent experimental data.

Carbon is one of the most cosmically abundant elements and C II is an
important ion in astrophysical sources such as the interstellar medium.
Study of accurate features of C II is of considerable interest and 
important for accurate spectral analysis. In their merged ion-photon 
beam experiment, Kjeldsen et al. \cite{kjl} measured the first 
detailed photoionization cross sections ($\sigma_{PI}$) of C II with 
high accuracy. With the synchrotron radiation from an undulator, their 
measurement exhibited highly resolved features of autoionizing 
resonances in the cross sections. The features agreed very well with 
the results from close coupling approximation using the R-matrix method 
\cite{n1}. However, the theoretical calculations did not include the 
relativistic effects and therefore did not predict the observed fine 
structure features.

The Opacity Project work did not include the fine structure.
The motivation for the present work is to study the fine structure
effects in relation to the observed features using the Breit-Pauli
R-matrix (BPRM) method \cite{st,ip,ip2}. The method has been extended 
to a self-consistent treatment of photoionization cross sections
and electron ion recombination (e.g \cite{znp}) and benchmarked for 
the total recombination cross sections in a unified treatment by 
comparing with measured recombination spectra \cite{znp,pnz}. The 
recombination cross sections require total contributions of 
photoionization cross sections of all bound levels. On the contrary, 
present work focuses on the features of photoionization cross sections 
of single levels of ground configurations, $2s^22p$, of C II, and 
compares with the recent measured cross sections.

The theoretical calculations for the photoionization cross sections
$(\sigma_{PI}$) are carried out in the close coupling (CC) approximation
using Breit-Pauli R-matrix method in intermediate coupling. 
Photoionization of the ion is described in terms of the eigenfunction 
expansion over coupled levels of the residual (`core' or `target') ion. 
The wavefunction of the (N+1)-electron ion is represented by the level
wavefunctions of the N-electron core multiplied by the wavefunction of 
the outer electron as follows:

\begin{equation}
\Psi(E) = A \sum_{i} \chi_{i}\theta_{i} + \sum_{j} c_{j} \Phi_{j},
\end{equation}

\noindent
$\chi_{i}$ is the target wavefunction in a specific level $J_i\pi_i$
and $\theta_{i}$ is the wavefunction for the ($N$+1)-th electron in a
channel labeled as $S_iL_i(J_i)\pi_ik_{i}^{2}\ell_i(\ J\pi)$; 
$k_{i}^{2}$ being its incident kinetic energy. $\Phi_j$'s are the 
correlation functions of the ($N$+1)-electron system that account for 
short range correlation and the orthogonality between the continuum 
and the bound orbitals. 

The BP Hamiltonian, as employed in the IP work \cite{ip}, is
\begin{equation}
H_{N+1}^{\rm BP}=H_{N+1}+H_{N+1}^{\rm mass} + H_{N+1}^{\rm Dar}
+ H_{N+1}^{\rm so},
\end{equation}
where $H_{N+1}$ is the nonrelativistic Hamiltonian,
\begin{equation}
H_{N+1} = \sum_{i=1}\sp{N+1}\left\{-\nabla_i\sp 2 - \frac{2Z}{r_i}
        + \sum_{j>i}\sp{N+1} \frac{2}{r_{ij}}\right\},
\end{equation}
and the additional terms are the one-body mass correction term,
the Darwin term and the spin-orbit interaction term respectively.

Present wavefunction for C II is expressed by an 20-level expansion 
of the core ion, C III, with configurations, $2s^2$, $2s2p$,
$2p^2$, $2s3s$, $2s3p$, $2s3d$, while the K-shell remains closed (Table I).
The core wavefunction was obtained from atomic structure calculations
with Thomas-Fermi potential using the code SUPERSTRUCTURE \cite{ss}. The 
spectroscopic and correlation configurations and the scaling parameters 
in the Thomas-Fermi potential are given in Table I. The correlation term 
in Eq. (1) considers all possible (N+1)-electron configurations formed 
from the maximum occupancies in the orbitals as $2p^3$, $3s^2$, $3p^2$, 
$3d^2$, $4s^2$, $4p^2$. 

The computations of $\sigma_{PI}$ are carried out using the package of 
BPRM codes \cite{ip2} from the Iron Project. The cross sections
are computed with a very fine energy mesh in order to delineate the 
detailed resonance structures as observed in the experiments. It is 
a computationally demanding procedure because of repeated computations 
with exceedingly fine energy bins to search and resolve the resonances.
Many resonances are sharp and narrow, and can not be detected 
individually in the experiment. They are convolved with a Gaussian 
function of FWHM equal to the energy bandwidth of the monochrometer of 
the experiment.

The calculated photoionization cross sections ($\sigma_{PI}$) of 
levels $^2P^o_{1/2}$ and $^2P^o_{3/2}$ of ground configuration 
$2s^22p$ of C II are presented in Fig. 1. Inclusion of relativistic 
effects in the BPRM approximation introduces more Rydberg series of 
resonances in the cross sections belonging to the increased number of 
core thresholds by fine structure splitting.

The BPRM cross sections are presented in the bottom panel of Fig. 1 
where $\sigma_{PI}$ of $^2P^o_{1/2}$ level is shown as the solid curve,
and of $^2P^o_{3/2}$ level as the dotted curve. Both cross sections have 
many overlapping resonances. They are compared with the photoionization 
cross sections of the $^2P^o$ ground state of C II in LS coupling (Fig. 
1a, upper panel). The 12 C III core states in LS coupling correspond to 
20 fine structure levels. The additional thresholds as well as 
resonances that are not allowed in LS coupling but are allowed in 
intermediate coupling (IC) have resulted in more extensive resonances
in the BPRM $\sigma_{PI}$. For example, photoionization of 
$2s^22p(^2P^o)$ is allowed to $^2S$ and $^2D$ channels only, not to 
$^2P$ in LS coupling. However, with the IC the (e+ion) symmetry $^2P$ 
is enabled since $^2P_{1/2}$ can mix with $^2S_{1/2}$, and $^2P_{3/2}$ 
can mix with $^2D_{3/2}$. The first series 
of resonances $2s2p(^3P^o)np(^2P)$ with $n=4,5,...$ (denoted as R1 series 
in Fig. 1b) appearing in BPRM calculations are not seen in LS coupling 
(Fig. 1a). The next two series of resonances, $2s2p(^3P^o)np(^2D,^2S)$ 
(denoted as R2 and R3 series) are common to both LS and BPRM. The 4th 
series of resonances, which could be identified as the combination of 
$2s2p(^1P^o)np(^2D,^2P,^2S)$, are also common (except for the 
$^2P$ component) although they are very weak in LS coupling calculations 
partially due to coarser energy mesh. Identification of this series differs 
from Kjeldsen et al. \cite{kjl} who assigned a different series with 
the same identification. This series was not observed in their 
measurement. Additional series of resonances are introduced starting
from the fourth complex which overlap with the others. Small shift in 
energy positions of the resonances, seen in the upper and lower panels 
of Fig. 1, is due to statistical averaging of LS energy terms over 
their fine structure components.

The BPRM photoionization cross sections are compared with the
measured cross sections by Kjeldson et al. (1999) after convolving the 
calculated resonances with a Gaussian distribution function with a FWHM 
of 35 meV, the monochrometer bandwidth of the experiment. The 
calculated ionization energy for the $^2P^o_{1/2}$ ground level is
1.7885 eV compared to the measured energy of 1.7921 eV, and for 
$^2P^o_{3/2}$ level it is 1.78793 eV compared to the measured one of 
1.7916 eV. The calculated cross sections have been slightly shifted to 
the measured ionization thresholds to compare with the experimenally 
measured $\sigma_{PI}$ in Fig. 2.

Fig 2a, presents the total detailed calculated cross sections of the 
levels $^2P^o_{1/2,3/2}$ (solid and dotted curves respectively) while  
Fig. 2b presents the convolved $\sigma_{PI}$ on the eV scale. The 
measured cross sections \cite{kjl} are presented in Fig. 2c. 
Very good agreement is seen in general in resonance features between 
the calculations (Fig. 2b) and the measurements (Fig. 2c). The resonance 
peaks are also in good agreement in the first three resonance complexes. 
However, some differences may be noticed for the higher complexes. The 
peaks of some calculated resonances at higher energies
are higher than the measured ones. These peaks are not expected to be 
damped by dielectronic recombination or the radiative decays of the
excited core thresholds; the decay rates ($\approx ~ 10^8,10^9~sec^{-1}$) 
are several orders of magnitude lower than typical autoionization rate
of $10^{13-14}~sec^{-1}$. These peaks were checked with inclusion of 
radiation damping effect with no significant reduction. However, several 
other reasons can explain the differences. The measured convolved cross
sections are not usually purely Gaussian as assumed in the calculated
cross sections. The same FWHM may not remain constant with energies and
the peaks may be lowered with a larger bandwidth. The detailed resonances 
can be convolved with varied bandwidths to conform to the observed 
features. Also, the cross sections might be further resolved with a finer 
energy mesh to reduce the differences in the high energy peaks. It may 
be mentioned that as resonances get narrower with energy approaching to the 
convergent threshold, the monochrometer requires a narrower bandwidth 
for finer resolution. To see the structures at higher energies near the
highest threshold, the resonances beyond 30.2 eV are convolved with 
a narrower FWHM, and hence seem sharper than the measured cross sections.  
Fig. 2d presents the convolved cross sections of excited $2s2p^2(^4P_J)$ 
levels in a smaller resonant energy region, from 24.2 eV to about 25.5 eV, 
as discussed below.

Photoionization cross sections of the three fine structure levels, $J=$
1/2, 3/2, and 5/2, of the first excited state $2s2p^2(^4P)$ of C II are 
also presented. Kjeldsen et al. \cite{kjl} found a few resonances of 
$2s2pnd$ (n=7,8,..) series from photoionization of $^4P$ state in the 
near threshold region of ground state of C II. 
These resonances belong to the allowed transitions in IC  among levels 
of $2s2p^2(^4P_J)$ and $2s2pnd(^2P^o_J,^2D^o_J)$, that are forbidden 
in LS coupling. Fig. 3a (top panel) presenting $\sigma_{PI}$ of $^4P$ 
state in LS coupling, shows that there are no resonances below the 
threshold at 1.872 Ry photoionizing to $2s2p(^3P^o)$ state of C III. 
However, extensive narrow resonances can be seen below this threshold with 
almost no background in the photoionization cross sections
of the fine structure levels $^4P_{1/2,3/2,5/2}$, shown separately as 
solid, dotted and dashed curves in Fig. 3. The ionization thresholds 
for these levels lie at about 1.4 Ry (pointed by arrow in the figure)
and each level exhibits presence of a near threshold resonance. As we 
are interested in the overlapping energy region with the ground state 
of C II, the resonances of the levels $J=$1/2,3/2,5/2 are resolved with 
a much finer energy mesh from $\approx$ 1.75 Ry to about 1.87 Ry. These 
narrow resonances are convolved with the same monochrometer bandwidth 
of 35 meV and are presented in Fig. 2d for comparison with the measured 
cross sections. The first three of the convolved peaks were identified by 
Kjeldsen et al. \cite{kjl} as they were observed in the experiment 
(Fig. 2c), indicating that there was a mixture of states $^2P^o$ and 
$^4P$ in their C II beam in the low energy region. The convolved 
resonance peaks are lower, except for the first one, than the observed 
ones. Further resolution of the resonances could conceivably have 
improved the agreement with the measured ones.

We have demonstrated that the theoretical fine structure photoionization
cross sections account for nearly all experimentally observed features.
The relativistic BPRM photoionization cross sections of the ground state 
and the first excited state of C II reveal the fine structure observed 
in experiments, but not obtained in the non-relativistic LS coupling 
calculation. Very good agreement is found between the BPRM cross 
sections and the measured cross sections \cite{kjl}. Convolution of the 
detailed resonances with a monochrometer bandwidth may result in some 
variations of peaks depending on the bandwidth, choice of energy 
distribution function and resolution of resonances.

Similar to the experimental work by Kjeldsen et al. \cite{kjl}, Phaneuf 
et al. \cite{petal} are currently measuring photoionization cross 
sections of C II at a different set-up at ALS (Advance Light Source). 
However, the resolution at ALS is much higher (the estimated 
monochromitc bandwidth is about 7 meV compared to 35 meV by Kjeldsen 
et al.). The cross sections from the present work are available 
electronically for comparison with other experiments.

This work was partially supported by the National Science Foundation
and the NASA Astrophysical Theory Program. The computational work was
carried out at the Ohio Supercomputer Center.

\def\amp{{Adv. At. Molec. Phys.}\ }
\def\apj{{ Astrophys. J.}\ }
\def\apjs{{Astrophys. J. Suppl. Ser.}\ }
\def\apjl{{Astrophys. J. (Letters)}\ }
\def\aj{{Astron. J.}\ }
\def\aa{{Astron. Astrophys.}\ }
\def\aasup{{Astron. Astrophys. Suppl.}\ }
\def\adndt{{At. Data Nucl. Data Tables}\ }
\def\cpc{{Comput. Phys. Commun.}\ }
\def\jqsrt{{J. Quant. Spect. Radiat. Transfer}\ }
\def\jpb{{Journal Of Physics B}\ }
\def\pasp{{Pub. Astron. Soc. Pacific}\ }
\def\mn{{Mon. Not. R. Astr. Soc.}\ }
\def\pra{{Physical Review A}\ }
\def\prl{{Physical Review Letters}\ }
\def\zpds{{Z. Phys. D Suppl.}\ }
\def\adndt{Atomic Data And Nuclear Data Tables}

\begin{figure}
\psfig{figure=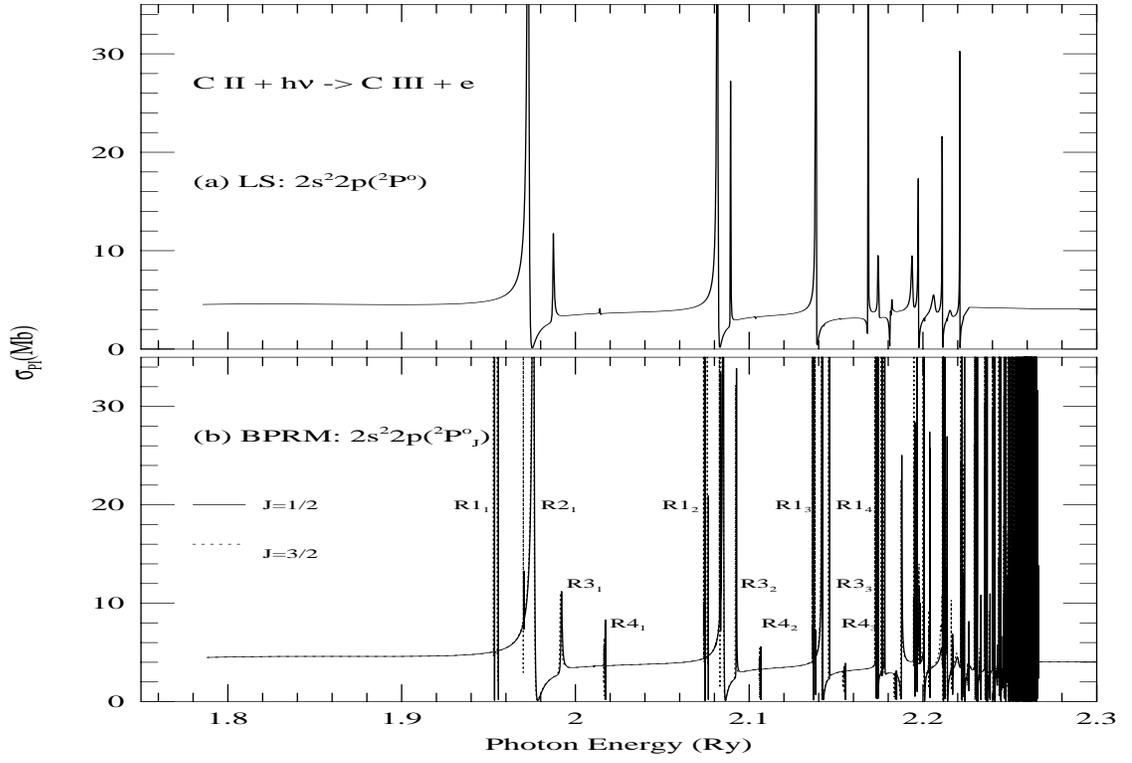,height=12.0cm,width=18.0cm}
\caption{Photoionization cross sections $\sigma_{PI}$ of the (a) ground 
$2s^22p(^2P^o)$ state in LS coupling, and (b) levels $^2P^o_J$, where 
$J$ =1/2 (solid curve) and 3/2 (dotted curve) in BPRM intermediate 
coupling, of ground configuration of C II. } 
\end{figure}

\begin{figure}
\centering
\psfig{figure=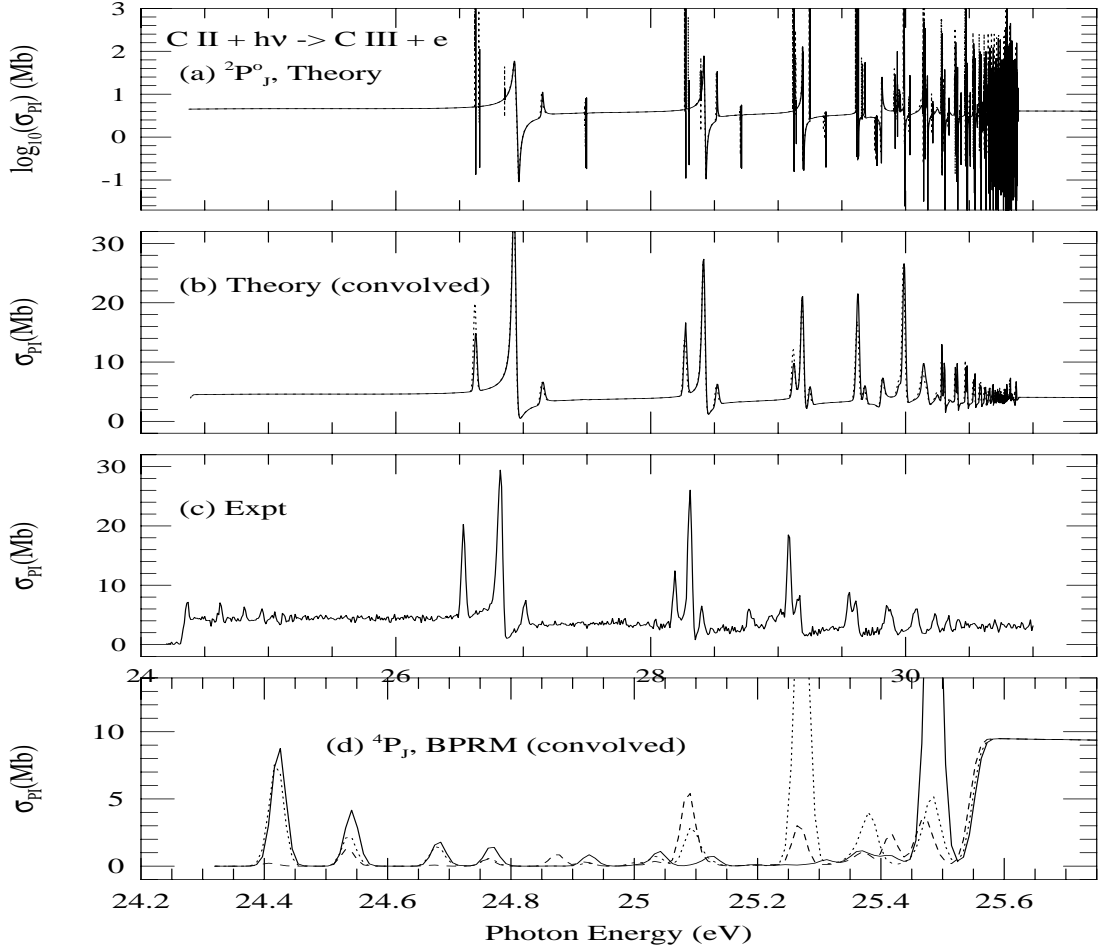,height=15.0cm,width=18.0cm}
\caption{Photoionization cross sections $\sigma_{PI}$ of C II: (a) 
detailed with resonances of levels $^2P^o_{1/2}$ (solid) and $^2P^o_{3/2}$ 
(dotted) of ground configuration $1s^22s^22p$, (b) the same cross sections
convolved with monochrormeter bandwidth of the experiment, (c) experimentally 
measured cross sections; (d) convolved cross sections of excited 
levels $2s2p^2(^4P_J)$, solid - J=1/2, dotted - J=3/2, dashed - J=5/2. }
\end{figure}

\begin{figure}
\centering
\psfig{figure=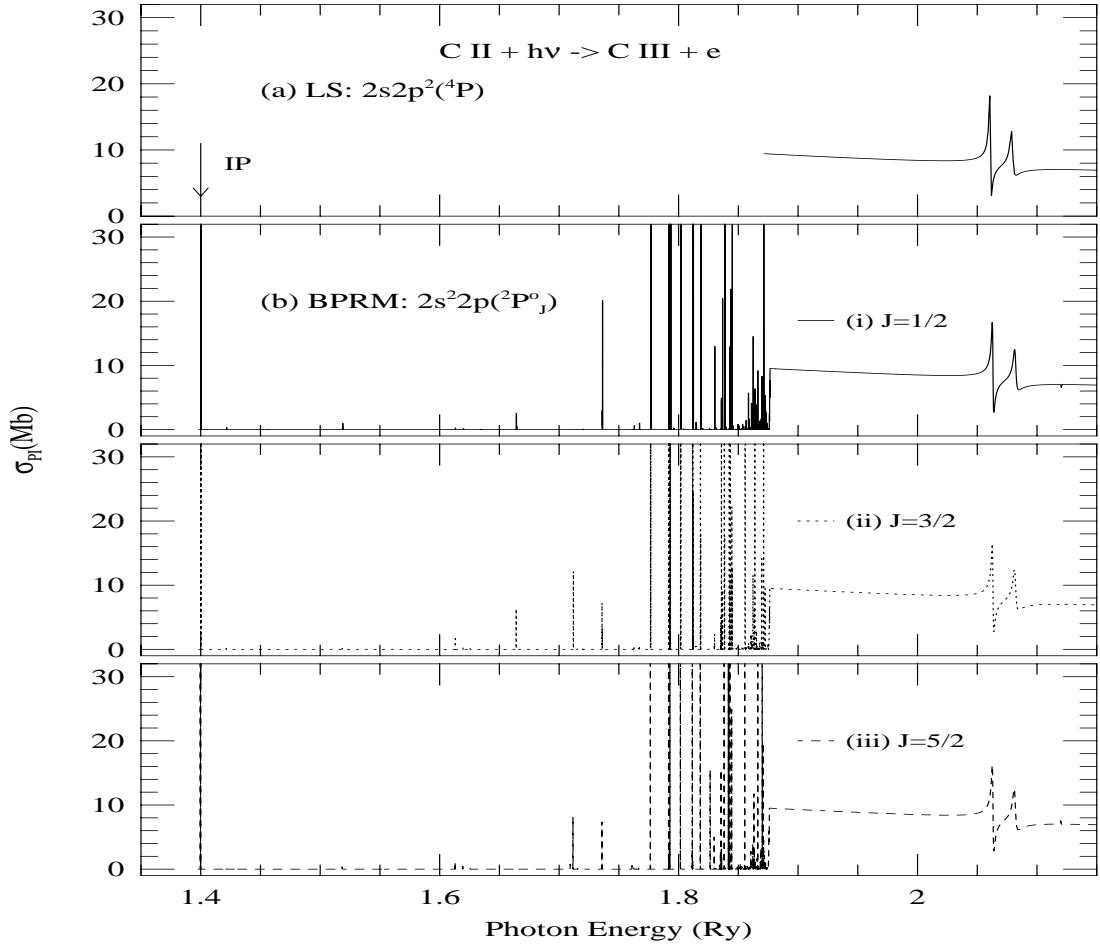,height=15.0cm,width=18.0cm}
\caption{Photoionization cross sections of the first excited 
$2s2p^2(^4P)$ state of C II: (a) in LS coupling, (b) of levels (i) J=1/2, 
(ii) J=3/2, and (iii) J=5/2 in BPRM approximation. Arrow points the 
ionization threshold for the fine structure levels.  }
\end{figure}

\narrowtext
\begin{table}
\caption{ Energy Levels of C III in the eigenfunction expansion of C II.
The list of spectroscopic and correlation configurations, and the
scaling parameter ($\lambda$) for each orbital are given below the table.
}
\begin{tabular}{rllc}
\hline
 & \multicolumn{2}{c}{Level} & \multicolumn{1}{c}{E(Ry)} \\
\hline
1 & $2s^2$ & $^1S_0$  &  0.        \\
2 & $2s2p$ & $^3P^o_2$  & 0.47793  \\
3 & $2s2p$ & $^3P^o_1$  & 0.4774   \\
4 & $2s2p$ & $^3P^o_0$  & 0.4772   \\
5 & $2s2p$ & $^1P^o_1$  & 0.9327   \\
6 & $2p^2$ & $^3P_2$    & 1.2530   \\
7 & $2p^2$ & $^3P_1$    & 1.2526   \\
8 & $2p^2$ & $^3P_0$    & 1.2523   \\
9 & $2p^2$ & $^1D_2$    & 1.3293   \\
10& $2p^2$ & $^1S_0$    & 1.6632   \\
11& $2s3s$ & $^3S_1$    & 2.1708   \\
12& $2s3s$ & $^1S_0$    & 2.2524   \\
13& $2s3p$ & $^1P^o_1$  & 2.3596   \\
14& $2s3p$ & $^3P^o_2$  & 2.3668   \\
15& $2s3p$ & $^3P^o_1$  & 2.3667   \\
16& $2s3p$ & $^3P^o_0$  & 2.3666   \\
17& $2s3d$ & $^3D_3$    & 2.4606   \\
18& $2s3d$ & $^3D_2$    & 2.4605   \\
19& $2s3d$ & $^3D_1$    & 2.4605   \\
20& $2s3d$ & $^1D_2$    & 2.5195   \\
\hline
\multicolumn{4}{l}{{\it Spectroscopic}: $2s^2$, $2s2p$, $2p^2$, $2s3s$,
$2s3p$, $2s3d$}\\
\multicolumn{4}{l}{{\it Correlation}: $2p3s$,$2p3p$,$2p3d$,$3s3p$,
$3s3d$,$2s4s$,$2s4p$,$4s4p$} \\
\multicolumn{4}{l}{$\lambda$: 1.42(1s),1.4(2s),1.125(2p),1.(3s),1(3p),1(3d),
3.3(4s),}\\
\multicolumn{4}{l}{3(4p)} \\
\end{tabular}
\end{table}


\end{document}